\documentclass[oldversion,letter]{aa} 
\usepackage{natbib}
\usepackage{graphicx}
\usepackage{txfonts}
\begin{document}
   \title{The nature of the soft X-ray source in DG~Tau}


   \author{P.C. Schneider
          \and
          J.H.M.M. Schmitt
          }


   \institute{Hamburger Sternwarte,
              Gojenbergsweg 112, 21029 Hamburg\\
              \email{cschneider/jschmitt@hs.uni-hamburg.de}
             }

   \date{Received .. / accepted ..}

  \abstract
   {The classical T Tauri star DG~Tau shows all typical signatures of X-ray activity and, in particular,  harbors a resolved X-ray jet. 
   DG~Tau's jet is one of the most well studied jets of young stellar objects, having been observed for more than 25 years by a variety of instruments.
   We demonstrate that its soft and hard X-ray components are separated spatially by approximately 0.2~arcsec by deriving the spatial offset between both components from the event centroids of the soft and hard photons 
   utilizing the intrinsic energy-resolution of the {\it Chandra} ACIS-S detector. We also demonstrate that this 
   offset is physical and cannot be
   attributed to an instrumental origin or to low counting statistics. 
   Furthermore, the location of the derived soft X-ray emission peak coincides with emission peaks observed for optical emission lines, suggesting that both, soft X-rays and optical emission, have the same physical origin.}

   \maketitle
%

\section{Introduction}\vspace{-2mm}
The evolution of protostars to young stellar objects (YSOs) is accompanied by accretion from a circumstellar 
envelope and disk as well as the loss of angular momentum by a substantial, often jet-like mass-outflow perpendicular to the disk \citep[e.g.][]{KP00}; however, neither the launching mechanism nor the collimation process leading to the observed jet-like outflows have been unambiguously identified.
The jet of DG~Tau is among the most well studied jets of YSOs, and these observations have placed tight constraints on the nature of the relevant processes, e.g. the detection of rotation in the outer regions of the DG~Tau jet \citep{Bacciotti2002} constrained the launch radius of the outflow.

DG~Tau is a classical T Tauri star (CTTS), whose basic properties were summarized by \citet{2007A&A...468..515G}.  
Its mass-outflow (a few $10^{-7}M_\odot\,$yr$^{-1}$ extending out to $\sim$~10\arcsec$\approx$2300~AU)
was first resolved by \citet{1983ApJ...274L..83M}. Most observations of DG~Tau's outflow were carried out in forbidden emission line regions (FELR), which trace material at temperatures of $\sim10^{4}\,$K and densities below $\sim10^{7}\,\mbox{cm}^{-3}$. These studies indicated that at distances larger than $\sim0.5$\arcsec~ from the central source the forbidden line emission is concentrated in individual blobs moving at velocities of 
$\sim300$~km/s (projected $\sim 0.3$\arcsec~yr$^{-1}$) approximately along the jet axis \citep{Pyo2003}.

The structure of the innermost region of the DG~Tau system is subject to permanent variations. Several studies
revealed evidence for material of different speeds and morphology in this region \citep[e.g.][]{Kepner93,Baciotti2000,Takami2002}, indicating an evolution on time scales of years;
as pointed out by \citet{1993ApJ...410L..31S}, the material in the vicinity of the star is probably denser than
in the more distant jet component.  In particular, the jet shows an onion-like structure, where the 
higher velocity material appears to be embedded in the more slowly moving material \citep{Bacciotti2002}.
The favored heating mechanism for jet emission is internal shocks, heating up the material to temperatures of $\sim 10^{4}$~K \citep{Lavalley2000}. 

As many (if not all) CTTS, DG~Tau also is an X-ray source, first detected by \citet{Feigelson81}. From the X-ray point of view the source is unusual in
two aspects. First, DG~Tau is the only stellar X-ray source harboring a resolved X-ray jet \citep{2008A&A...478..797G},  
which can be traced out to a distance of $\sim$~5\arcsec~ from the central source with a
luminosity of about 10\% of the central soft X-ray component. Second,
the X-ray properties of DG~Tau resemble that of the class of ``two-absorber-X-ray (TAX) sources'' \citep{2007A&A...468..515G,2008A&A...478..797G}. X-ray spectra of TAX sources are basically the sum of two thermal components, 
differing not only in mean temperature but - in contrast to most other X-ray spectra - also in absorbing column density.
In DG~Tau, the emission regions of the soft and the hard components appear to be disjoint spatially.  In an {\it XMM-Newton}
observation, \citet{2007A&A...468..515G} found an increase in the hard component's count rate during a flare, while the soft
component's count rate remained constant; they proposed therefore, supported by the spectral properties of the soft component, an interpretation of the soft component as internal shocks in a jet close to the star.
Motivated by these indications that the soft X-ray component in DG~Tau might be spatially detached from the hard X-ray component, we performed a detailed position analysis of both components utilizing the superb angular resolution of the {\it Chandra} telescope.

\vspace*{-5mm}
\section{Observations, data processing, and data analysis}
\vspace*{-1mm}
The available $Chandra$ data of DG~Tau cover a total exposure time of 90 ks split into 4 individual observations performed between 2004 and 2006 (see Table 1 of \citet{2008A&A...478..797G} for a summary). The details of these observations were presented by \citet{2008A&A...478..797G}.
Our data reduction was completed using CIAO Version 4.0, along the lines of the $Chandra$\, analysis threads with the aim to derive accurate source positions.  We define a soft ($0.3-1.1$~keV) and a hard ($1.7-7.0$~keV) spectral component and list their
relevant properties in Cols.~3 and 4 of Table~\ref{dgoffsets}; because of the TAX property of DG~Tau, the mutual 
contamination of the components is quite small.

The $Chandra$-calibration team states a 0.1\arcsec ($1\,\sigma$) accuracy for relative positions on the ACIS S3 detector \footnote{http://cxc.harvard.edu/cal/docs/cal\_present\_status.html\#rel\_spat\_pos}.  This implies that an offset between the central source
and X-ray emission arising in the inner part of the optically resolved jet is -- at least in principle -- measurable, given the
fact that the stellar emission component is thought to be strongly absorbed at soft X-ray energies below 1~keV. 
We, therefore, derived individual positions for the above defined soft and hard components and
calculated their respective centroids with sub-pixel resolution.  The most precise determination of source positions is
complicated by the fact that the superb point-spread function of the $Chandra$ mirrors is slightly undersampled by the ACIS-S detector.
To compensate for the effect of this undersampling, various strategies, such as sub-pixel event repositioning \citep{Li2004}, can be pursued during the data processing and data analysis. During pipeline processing, the nominal photon positions are randomly distributed within a given detector pixel ($\pm~0.25$\arcsec). Alternatively, no randomization \citep[cf.,][]{Feigelson2002} or re-randomization schemes can be applied.
We pursued a conservative approach and used the archival data with standard randomization to minimize possible aliasing effects, and verified that our results (and our conclusions) do not depend on the type of the randomization chosen.

To determine source positions, we experimented with the standard source-detection tools \texttt{wavdetect} and \texttt{celldetect}. The \texttt{celldetect} algorithm
uses different photons (the ``search region'' is always a box in the projected image which is not necessarily centered on the centroid) and the \texttt{wavdetect} algorithm reverts to binned data, thus both approaches are not optimized to find the most accurate source position.  Therefore we developed our own iterative source position determination algorithm.
Starting with an approximate ``by eye'' position, we extracted all photons within a 0.75\arcsec~ radius around this
position to determine a new centroid; with this new position photons were then reextracted and the entire process
continued until convergence.  This  method should operate well for a symmetric point response function, which applies in the central FOV.
We note that the soft component's size might actually deviate from that expected from a point-like source on the order of 0.5~\arcsec, while we can exclude a size of $\ge1$\arcsec~ with $>90$\% confidence. However, the signal-to-noise (SNR) of the available data does not enable statistically robust results to be derived.

 \begin{table}
\caption{Offsets for the individual observations} 
\label{dgoffsets}
\centering
\setlength\tabcolsep{4pt}
\begin{tabular}{c c c c c c c}
\hline\hline
Obs-ID & Offaxis &Soft  & Hard  & Offset & Position angle\\
& (arcmin) & photons & photons & (arcsec) &(degree) \\
\hline
   4487 & 1.43 & 138 & 191 & 0.23 & 225 \\
   6409 & 0.55 & 67 & 112 & 0.20 & 215 \\
   7247 & 0.55 & 65 & 49 & 0.13 & 191 \\
   7246 & 0.55 & 133 & 187 & 0.20 & 215 \\
\hline
\end{tabular}
\vspace*{-4mm}
\end{table}

The ``very faint''-mode of the observations leads to a low probability of finding a background photon in our search region,
which is below 30\% for the longer exposures. A single photon shifts the derived source position by less than 0.03\arcsec, so that any background is essentially negligible in our analysis.

During the analysis, we kept the 4 individual exposures separate. Merging the individual exposures can degrade the spatial resolution because the absolute astrometric accuracy of $Chandra$\, ($\sim$ 0.4\arcsec (1$\sigma$)) is worse than its relative accuracy. To account for this effect, we reprojected the individual exposures so that the centroid of the hard component of DG~Tau was aligned in all observations.
We coadded the images to be able to derive higher signal-to-noise data for a cross-check; we are aware, however, that
the positions of the soft X-ray emission might not be constant throughout the observation  period of almost two years.

\section{Results}
\subsection{Soft and hard source positions}
\begin{figure}
    \centering
   \includegraphics[width=0.49\textwidth]{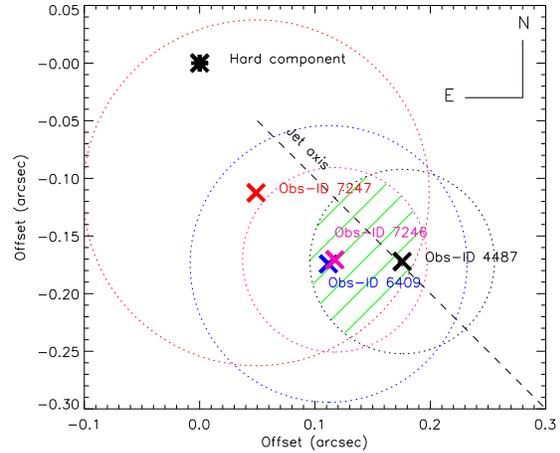}
   \caption{Relative spatial offset of the soft X-ray component. The hard component is centered at (0,0). The circles indicate the 90\% confidence interval taken from Table \ref{props} and Obs-ID 3730. The shaded area is included in the 90\% confidence ranges of all observations.}
\label{offsets}
\end{figure}

\vspace{-2mm}
In Fig.\ref{offsets}, we show the computed separations of the soft and the hard X-ray centroid for all four observations of DG~Tau 
and their estimated error radii. We note that all derived positions were shifted 
in order to align the position of the hard component in all observations.  As is clear from  Fig.\ref{offsets},
all observations exhibit a (sub-pixel) offset between the soft and the hard X-ray components.  For the three well exposed observations (cf., Table~\ref{dgoffsets}), we find similar offsets with a separation of $\sim0.2$\arcsec, while the fourth observation (Obs-ID 7247) shows a smaller offset, but in a similar direction, and - considering the low count statistics - still compatible with the other observations.
A total offset of 0.21\arcsec~ is also obtained if the centroids in the coadded event-files are considered, which equals the best-fit value derived by using all individual observations with the errors estimated in Sect.~\ref{notstat}.
The position angles of the measured DG~Tau offsets yield a best fit offset angle of $\sim$~218$^\circ$, which compares well
with the position angle of $\sim 225^\circ$ for the jet orientation in the optical \citep[e.g. ][]{Eis98}.

\vspace*{-5mm}

\subsection{Is it instrumental ? }
\begin{figure}
  \centering
  \includegraphics[width=0.49\textwidth,height=0.27\textwidth]{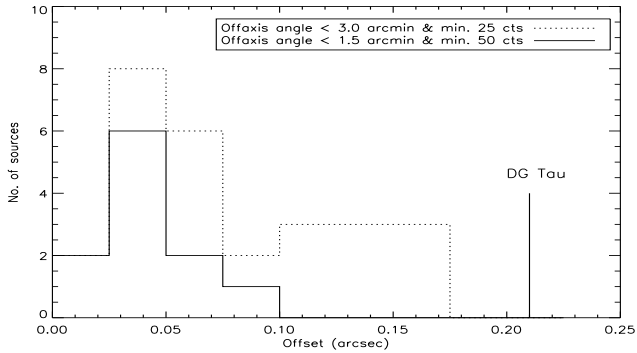}
  \caption{Statistics of the comparison observations.}\label{comp_obs}%
\end{figure}

\vspace{-2mm}
To investigate whether the observed offset between soft and hard source positions can be attributed to instrumental
effects, we retrieved a number of observations from the $Chandra$ archive taken with
ACIS-S3 in the VFAINT-mode (Obs-ID 3730 is FAINT-mode).  The retrieved observations are listed in Table~\ref{testobs}.  For these targets,
we then performed the same analysis as for the DG~Tau data and computed offsets between the soft and hard source positions. 
In Fig. \ref{comp_obs}, we show the derived offset statistics, where we distinguish between the ``good'' data sample
(off-axis angle $<\, 1.5$\arcmin~ and $>$ 50 cts) and the ``poor'' data sample (off-axis angle $<\, 3$\arcmin~ and $>$ 25 cts);
we note that in this nomenclature our DG~Tau is ``good'' data.  Figure \ref{comp_obs} clearly shows that DG~Tau's offset is extremely
unusual, and in fact none of the investigated data sets and in particular none of the ``good'' data sets shows an offset 
comparable to that observed in DG~Tau.

\vspace{-0mm}

We investigated the energy dependence of the point response function by calculating the centroids of precomputed synthetic PSF-images provided by the CIAO-tool \texttt{mkpsf} for the DG~Tau source position and
found that the centroid position changes by less than 1/100 pixel between the hardest (7 keV) and the softest photons (0.3 keV), even for Obs-ID 4487, which has the largest off-axes angle. Performing Marx 4.3 simulations \footnote{http://space.mit.edu/ASC/MARX/} that include a model of the individual mirror shells strengthened this finding.
Then we further restricted the energy ranges of the test-exposures with a high count-statistics such as Obs-ID~3730, 4470, 49899, and 626 to the outer edges of the energy bands, e.g. 0.3-0.7~keV and 3.0-7.0~keV, to check whether the offset becomes larger. A positive result would imply that the centroid position is dependent on the energy range used. However, the measured separations differ only slightly from the previous results, and remain, in any case, far smaller than our DG~Tau offsets.
Finally, we split the hard X-ray component of DG~Tau which represents the coronal emission in our interpretation into two groups. These offsets within the hard component are all compatible within a $\sim0.1$\arcsec~ margin of error (considering the lowered count statistics by splitting the hard photon group into two). The position angle also differs from that of the separation of the hard and the soft component.
In summary, we conclude that no evidence supports the idea that the measured offset can be attributed
to an instrumental effect.

\vspace*{-2mm}
\subsection{Is it statistical ?\label{notstat}}
\begin{figure}
  \centering
   \includegraphics[width=0.49\textwidth,height=0.27\textwidth]{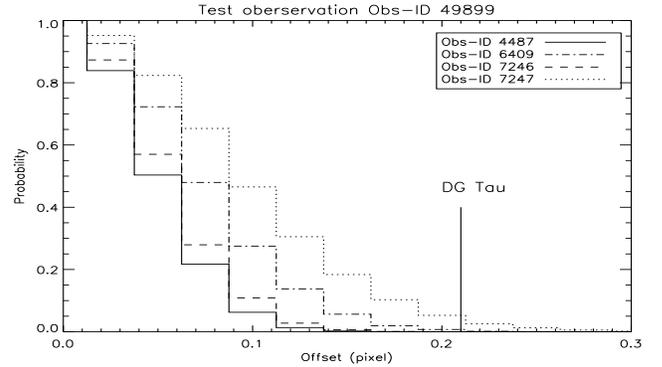}
   \caption{Distribution of distances between the soft and the hard X-ray component for point-like sources with simulated count statistics matching the circumstances of the DG~Tau observations.}\label{stat_offset}%
\end{figure}

\vspace*{-2mm}
We investigated the statistical errors in the soft and hard photon centroid positions.  To assess 
the statistical scatter in the source positions, we artificially reduced the number of photons 
in the high count statistics test  observations (cf., Table \ref{testobs}) by selecting randomly the same number of
``source'' photons as observed in the individual DG~Tau observations in the desired energy range, and
computed the source centroid positions and their offsets.
An example of the simulated offset distribution (using $10^4$ realizations) is shown in Fig. \ref{stat_offset}
and the relevant properties of the distributions are summarized in Table \ref{props}. 
As is obvious from Fig. \ref{stat_offset} and the numbers in Table \ref{props}, statistical fluctuations are an 
extremely unlikely cause of the observed DG~Tau offset.

\begin{table}
\caption{Probabilities derived from the comparison observations. The 90\% limit refers to the seperation that only 10\% of the trials exceed and the probability is the derived value of finding the measured or a larger offset.}\label{props}
\centering
\begin{tabular}{c c c c}
\hline\hline
Comparison & DG~Tau & $d<$ for & Probability\\
observation & observation & 90 \%& \% \\
\hline
3730 & 4487 & 0.08" & $<$0.1\\
     & 6409 & 0.12" & 0.4\\
     & 7246 & 0.08" & $<$0.1\\
     & 7247 & 0.15" & 21.2\\
4470 & 4487 & 0.11" & $<$0.1 \\
     & 6409 & 0.15" & 1.2\\
     & 7246 & 0.12" & $<$0.1\\
     & 7247 & 0.19" & 32.3 \\
\hline
\end{tabular}
\vspace*{-2mm}
\end{table}

\subsection{Offset significance and uncertainties for DG~Tau}
\vspace*{-2mm}
Neither the studied comparison sources nor our simulations show an offset between the soft and the hard photon centroids 
sufficiently large to explain the observed offsets in DG~Tau.
The probability of measuring an offset larger than 0.2\arcsec~ is below  $\sim 0.01$ for the observations of longer exposure times (Obs-OD 4487, 6409, 7246), if both sources are at the same position.  We note that the measured position angle between soft and hard position correlates 
with the optically known jet-direction. Using an estimate of the measured position angle distribution of $\pm 30^\circ$ about the jet-direction (cf., Table~\ref{dgoffsets}), the probability that the measured position angle is located in the same range for all observations is only $7.7\times10^{-4}$, assuming that they are distributed uniformly.  Thus, formally, we estimate a probability of less than $10^{-8}$ that the
observed offset distribution is obtained by chance and therefore conclude that any systematic errors are far smaller than the observed offsets and that
the errors in our measurements are dominated by counting statistics.

\begin{table}
\caption{Comparison observations}\label{testobs}
\centering
\tabcolsep1mm
\begin{tabular}{c l c c c}
\hline\hline
Obs-ID & Source & Off-axis & Min.  & Offset\\
& & (') & counts & (") \\
\hline
4470 & Gl 569 A & 0.5 & 1565 & 0.03 \\
49899 & Prox Cen & 0.6 & 1040 & 0.02 \\
971 & TWA-5 &  0.3 & 568 & 0.03\\
3730 & GJ 3275 & 0.6 & 492  & 0.02\\
6416 & NGC 1977 311 & 2.9 & 436 & 0.04 \\
626 & HD 113703 B & 0.7 & 400 & 0.05 \\
6417 &  NGC 1977 311 & 2.8 & 205 & 0.07 \\
4476 & GJ1245 A &  0.5 & 118 & 0.03\\
626 & HD 113703 C          & 0.7 & 99 & 0.08 \\
4489 & 2MASS 05352360-0628244 & 2.63& 94 & 0.11\\
6417 &  CSV 6218         & 2.8 & 88 & 0.17\\
4510 & NGC6791 KU B16&   0.9 &  86 & 0.05 \\
6416 & V* V372 Ori  & 0.7 & 81 & 0.04 \\
4476 & GJ1245 B &  0.5 & 63 & 0.04 \\
4485 & 1WGA J2203.9-5647 & 1.7 & 56 & 0.16\\
5427 & HD 179949 & 0.3 & 49 & 0.08\\
627 & HD 129791B & 1.7 & 45 & 0.13 \\
4510 & NGC6791 SBG 9315 & 1.4 & 44 & 0.11 \\
6121 & HD 179949 & 0.6 & 43 & 0.06 \\
7247 & 2MASSs J0426573+260628 & 1.7 & 41 & 0.07 \\
630 & CXOSEXSI J175823.5+663950 & 1.5 & 36 & 0.16 \\
6417 &    JW 94       & 1.5 &  34 & 0.06 \\
6120 & HD 179949 & 0.3 & 31 & 0.04\\
4488 & FS Tau & 0.8 & 31 & 0.14\\
4487 & 2MASSs J0426573+260628 & 1.9 & 31 & 0.13 \\
6417 & V* V372 Ori & 0.8 & 29 & 0.12\\
6416 & Parenago 1606 &  1.4 & 25 & 0.08 \\
\hline
\end{tabular}
\vspace*{-4mm}
\end{table}

\vspace*{-4mm}
\section{Discussion}
\vspace*{-2mm}
By considering the measured offset of 0.2\arcsec~ between the soft and the hard X-ray centroid position as physical, we can convert this offset into
a physical distance of 48~AU from the central source, assuming a distance of 140~pc and a disk
inclination of $38^\circ\,$\citep{Eis98}.  This distance is an order of magnitude larger than reasonable launching regions of the jet \citep[$\sim$~1~AU,\,][]{Anderson2003}, and our results therefore favor strongly the interpretation that the X-ray emission observed close to the star is originating from internal shocks of the jet, as proposed by \citet{2007A&A...468..515G}. Internal shocks are incidentally also the preferred heating mechanism for the optically observed FELRs. With this interpretation, the total X-ray
luminosity of the jet is an order of magnitude higher than that of the $Chandra$-resolved part of the jet \citep{2008A&A...478..797G}, although even then its luminosity is far smaller than the optical jet luminosity; for example, \citet{Lavalley2000} derived
a luminosity, for the first emission peak of the [O~I]-line, of $1.1\times 10^{-4} L_\odot$, which is a factor of
$\sim40$ higher than the energy-loss by X-ray emission, suggesting that only a small amount of the outflowing material
reaches X-ray temperatures.
We now consider the following scenario:  we model the X-ray jet by a cylinder of radius $r$ and height $d$;
the base of the cylinder is located at the shock region, where the material is heated to some temperature $T$, and the shocked material flows through this cylinder with some post-shock velocity $v$.
We assume that the shocked material cools predominantly by radiation with a cooling time $\tau_c$ given by
$\tau_c = 3k_B T / (n \Lambda(T))$,
where $n$ is the plasma density, $T$ the (post-shock) temperature, $\Lambda(T)$ the cooling function, and $k_B$ Boltzmann's constant.
The cooling distance $d$, i.e., the height of the cylinder, is then given by
$d = \tau_c \cdot v = 3k_B T v/(n\cdot\Lambda(T)) $.
We further know the total emission measure $EM$ of the X-ray emitting plasma and write
\begin{equation}
EM =  f\cdot n^{2} \cdot A\cdot d = 3 f\cdot n\cdot\pi r^{2} \cdot v k_B T/\Lambda(T),
\end{equation}
where $f$ denotes an unknown filling factor of the hot plasma.
The mass outflow rate $\dot{M}_{X-ray}$ of the X-ray emitting plasma can be computed from 
\begin{equation}
\dot{M}_{X-ray} = m_H \ f \ n\cdot \pi r^{2} \cdot v= m_H\Lambda(T) EM/(3 k_B T),
\end{equation} 
i.e. the mass outflow rate of the X-ray emitting material is only determined by the observed
quantities $T$ and $EM$.
Our spectral fit provides a mean temperature of $T\sim3.4$~MK for the shocked plasma and \mbox{$EM \sim 3.5\times10^{52}\,\mbox{cm}^{-3}$} (APEC-models, metallicity at 0.3~solar) compatible with the values given by \citet{2008A&A...478..797G}. With these numbers we find an outflow rate of 
$1.3\times 10^{-11} M_\odot/\mbox{yr}$
using 
\mbox{$\Lambda(T)\approx2\times10^{-23}~\mbox{erg cm}^{3}\mbox{s}^{-1}$}. This value is indeed orders of magnitude smaller than the outflow rate of the high velocity material only \citep[$4\times 10^{-9} M_\odot/\mbox{yr}$, ][]{Lavalley2000}.
Are such values physically reasonable? We note that the soft X-ray component is more or less point-like if we disregard
for the time being the extended jet component described by \citet{2008A&A...478..797G}.  Given the distance of 140~pc towards DG~Tau, this implies that the region $d_{max}$ has the approximate size of 112~AU (one ACIS pixel). If we estimate the outflow speed of the shocked material to be approximately 300~km/s, which is on the one hand the speed of the high-velocity material measured in FELRs and
on the other hand approximately the speed required to produce the observed soft X-ray temperature by means of the strong shock formula \citep[the optically observed shocks have speeds of only up to 100~km/s,][]{Lavalley2000}
we can then derive
 $n > n_{min} =  3k_B T v/\left(d_{max}\cdot\Lambda(T)\right) = 1.3 \times 10^{6} \mbox{cm}^{-3}$.\\
Using these values of $n_{min}$ and $d_{max}$ to calculate the emission measure, we estimate the effective outflow cross sectional
area \mbox{$f \pi r^{2} \approx 1 \times 10^{25}$ cm$^{2}$} or \mbox{$f \pi r^{2} \approx 6 \times 10^{-2}$ AU$^2$}. 
The launch and collimation distance of the jet in DG~Tau is believed usually to be approximately \mbox{1 AU}.  Therefore, the filling factor of the X-ray emitting material must be small and we envisage a scenario of hot X-ray plasma with a small filling factor, immersed in cooler material with a far larger filling factor.
A fraction of the shocked material is clearly observed radiating in the resolved jet at a distance of 5\arcsec~ to DG Tau at essentially the same X-ray temperature as the ``inner jet''. The cooling time of this material may be sufficiently large to enable the material to move the distance required; this would descrease the density and the filling factor by one and two orders of magnitude, respectively, in comparison to the values above. We prefer, however, an interpretation in which the resolved jet is possibly ``re-shocked'' material. At any rate, only a minor fraction of the outflowing material in the ``inner jet'' experiences shocks at $\sim 300$~km/s or more, while 
the densities and mean velocities of the hot (T $\sim 3.4\times 10^{6}$~K) and the
cool (T $\sim 10^{4}$~K) material are similar to within factors of a few.

\section{Summary}
\vspace*{-2mm}
Our detailed analysis of the spatial distribution of the X-ray emission of DG~Tau shows that
the soft and the hard X-ray emission can be spatially separated, which is consistent with the suggestion of \citet{2007A&A...468..515G}.
The measured separation is 0.21"$\approx48$~AU and the X-ray jet of DG~Tau is therefore not only located at large distances 
(up to 5\arcsec) but also close to the stellar emission.
If we identify the hard X-ray component with coronal emission from the stellar surface, which is suggested by 
its stronger absorption compared to the soft component, then the position of the soft X-rays coincides with a region in the 
DG~Tau jet, where enhanced emission in the FELRs is observed.
Only a small fraction of the total mass-loss and the radiative loss is needed to explain the observed X-rays, and therefore
only a small fraction of the outflowing material appears to reach X-ray emitting temperatures. Unfortunately, the available observations do not allow any detailed studies of the spectral features of the soft X-ray component, and therefore a grating observation of DG~Tau would provide deeper insights into the true nature of the soft X-ray component's emission process.


\begin{acknowledgements}
      This work has made use of data obtained from the $Chandra$ data archive.
      P.C.S. acknowledges support from the DLR under grant 50OR703.
\end{acknowledgements}
\vspace*{-2mm}
\bibliographystyle{aa}
\vspace*{-4.5mm}
\bibliography{dg_main}
\end{document}